\documentstyle[aps,preprint,epsfig]{revtex}
\newcommand{\beq}{\begin{equation}}
\newcommand{\eeq}{\end{equation}}

\begin{document}
\draft{}

\title{Inhomogeneous Multidimensional Cosmologies}
\author{ Santiago E. Perez Bergliaffa\footnote{E-mail: santiago@lafexsu1.lafex.cbpf.br}}
\address{Centro Brasileiro de Pesquisas F\'{\i}sicas\\
  Rua Xavier Sigaud 150, CEP 22290-180, Rio de Janeiro \\
Brazil}
\maketitle
\begin{center}
\end{center}
\begin{abstract}
Einstein's equations for a 4+$n$-dimensional inhomogeneous space-time are presented, 
and a special family of solutions is exhibited for an arbitrary $n$. The solutions 
depend on two arbitrary functions of time. The time development
of a particular member of this family is studied. This solution 
exhibits a singularity at $t=0$
and dynamical 
compactification of the $n$ dimensions. It is shown that the behaviour of the system in the 4-dimensional ({\em i.e.} post-compactification) phase is constrained by the way in which the compactified dimensions are stabilized. The fluid that generates this solution is analyzed by means of the energy conditions.
\end{abstract}

PACS number(s): 04.50.+h     98.80.-k    11.25.Mj
\vspace{1.5cm}


\section{Introduction}

Over the last two decades, increasing attention has been paid to theories that 
unify the fundamental interactions in more than three spatial dimensions. The 
story of this kind of theories started in the 20's, when Kaluza \cite{kal} and
Klein 
\cite{kle} augmented the dimensionality of space to describe both gravity and 
electromagnetism as manifestations of geometry, using the degrees of 
freedom available from the 5-dimensional metric tensor \cite{bl}. The idea was 
renewed in the 60's by
deWit \cite{dewit} who tried to incorporate non-abelian interactions into the 
scheme. 
The original idea of Kaluza and Klein turned out to be 
incomplete for several reasons, 
but it still
pervades in one way or another many of the unifying schemes currently 
thought to be viable 
(most notably in the case of string theory. See for instance \cite{still}). 
However, if we are willing to accept any of these theories in which
space has more than three dimensions, we 
are faced with several 
questions, particularly on the cosmological side. Perhaps the most obvious one is related to 
the fact that 
we live in a 4-dimensional space-time, so every theory formulated in more than 
4 dimensions 
must say something about the fate of the extra dimensions. A convenient working
hypothesis would be to assume 
that they have been compactified up to some small size.
From a theoretical point of view, the most satisfactory way to achieve 
the compactification of the extra dimensions would be
the dynamical one. This means that the theory has solutions in which
the size of the extra dimensions diminishes as the universe evolves.
Solutions of this type have been found for the more symmetric cases
both in 4+1 and 4+$n$ dimensions \cite{bl}, but only a few
with some degree of inhomogeneity can be found in 
the literature, and always for the 4+1 case (see \cite{hin1} 
and references therein). 
Here instead a $4+n$ dimensional model with arbitrary $n$ will be studied. 
This case may have a paramount importance, as shown by the recent work of
Arkani-Hamed {\em et al} \cite{arkani}, in which the existence of $n$ sub-millimeter 
dimensions (with $n\geq 2$) yields a new framework for solving the hierarchy problem,
which does not rely on supersymmetry or technicolor. The central idea of this 
scheme is
that the existence of these extra dimensions
brings quantum gravity to the Tev scale through the
relationship between the Planck scales of the $4+n$ dimensional 
theory and the long-distance 4-dimensional theory. It must be remarked that
the extra dimensions are supposed 
to have a characteristic length of less than a millimeter, in accordance with
the lower bound at which gravity has been tested up to date \cite{long}.
In this framework, 
the fields of the Standard Model are localised on a 
3-brane in the higher dimensional space. Some of
the important consequences of these ideas in 
phenomenology, astrophysics, and cosmology can be found in \cite{arkanirev}. 
Many papers related to these matters have appeared lately; we mention here only a few. 
Argyres {\em et al} 
\cite{argyres} have studied the properties of black holes with Schwarszchild
radius smaller than the size of the extra dimensions, and 
concluded that the spectrum of primordial black holes in a 4+n dimensional 
spacetime differs from the usual one. Moreover, these 
primordial black holes would provide dark matter candidates and seeds for 
early galaxy and QSO formation. 
Mirabelli {\em et al} \cite{mira} have recently analyzed the missing-energy 
signatures that should be present in high-energy particle collisions due to the 
radiation of gravitons if gravity is important at TeV scale. They argue that collision
experiments provide the strongest present constraint on the size of the extra 
dimensions. 
Nath and Yamaguchi \cite{nath} have explored the effect of the excitations associated 
with extra dimensions on the Fermi constant. They give stringent constraints on the
compactification radius from current precision determinations of the Fermi constant, of the 
fine structure constant, an of the mass of the W and the Z bosons. 

A salient feature of the model we propose here is its inhomogeneity. 
This type of models might
describe an early phase of the universe, or may be of use on a super-horizon
scale, as suggested by chaotic inflation \cite{caoin}.
Besides, the work of Mustapha {\em et al} \cite{ellis} indicates that
there is no unquestionable observational evidence for spatial homogeneity.
This makes worthwhile analysing models that are isotropic but exhibit
some degree of inhomogeneity \cite{kra}. 

The aim of this paper is then to show the existence
of analytical solutions in inhomogeneous cosmological models 
in $4+n$ dimensions. Although some exact solutions for the 4+1 dimensional 
inhomogeneous case have been worked out \cite{hin1}\cite{hindues}, the case 
dealt with here has not been
studied previously. Due to the complexity of the equations of motion (which have not been
displayed before in the literature in the case of an arbitrary $n$), the 
inhomogeneity has been restricted to the $n$ internal dimensions. 
It will be shown here that there exist solutions for which the 4-dimensional spacetime is expanding while the extra dimensions conpactify due to the evolution of the system.
Also, some remarks will be made on the dependence of the evolution of the system after the compactification on the stabilization of the extra dimensions. Finally, the matter content of the system in the multidimensional phase will be characterized by the study of the strong and weak energy conditions (SEC and WEC respectively).

\section{Field equations and solutions}

The starting point is the $4+n$ dimensional metric, given by
\beq
ds^2=-dt^2+e^{2\lambda (t,r)} (dr^2+r^2\;d\Omega ^2)+e^{2\mu(t,r)} dy^2 ,
\label{metric}
\eeq
where $d\Omega ^2$ is the surface element on the 2-sphere, and 
$dy^2\equiv\sum^n_{i=4} dy_i^2$. For simplicity we will work with a plane 3-space,
and we assume a single scale factor for the internal dimensions.

We adopt the following stress-energy tensor for the matter content of the model:
\begin{equation}
T_{\mu\nu} = {\rm diag} (\rho , -p_3, -p_3, -p_3, -p_n,...,-p_n),
\label{set}
\end{equation}
with $p_n$ the internal pressure. The nonvanishing field equations in this $4+n$ dimensional 
space-time are then
\begin{mathletters}

\begin{equation}
-2\dot\lambda '-n\dot\mu '+n\dot\lambda \mu ' -n\dot\mu \mu ' =0 ,
\label{eeq1}
\end{equation}

\beq
2\ddot\lambda + 3\dot\lambda^2 + 2n\dot\lambda\dot\mu+n\ddot\mu+
\frac{n(n+1)}{2}\dot\mu^2-e^{-2\lambda}\left[ \frac{2\lambda'}{r}+\frac{2n}{r}\mu ' +2n
\lambda ' \mu '  +\lambda '^2-\frac{n(1-n)}{2}\mu '^2\right]  = -8\pi p_3 ,
\label{eeq2}
\eeq

\begin{equation}
2\ddot\lambda + 3\dot\lambda^2 + 2n\dot\lambda\dot\mu+n\ddot\mu+
\frac{n(n+1)}{2}\dot\mu^2-e^{-2\lambda}\left[\lambda ''+\frac{\lambda '}{r} + 
n\mu '' + \frac{n}{r}\mu ' +\frac{n(n+1)}{2}\mu '^2\right]=-8\pi p_3 ,
\label{eeq3}
\end{equation}

\begin{eqnarray}
(n-1)\ddot\mu+\frac{n(n-1)}{2}\dot\mu ^2+3(n-1)\dot\lambda\dot\mu +
3\ddot\lambda+6\dot\lambda ^2 - e^{-2\lambda}\left[ 
(n-1)\lambda ' \mu ' + \right. & &  \nonumber \\
\left. (n-1)\mu '' +\frac{n(n-1)}{2}\mu '^2+\frac{2(n-1)}{r}\mu ' +2 \lambda '' + 
\frac{4}{r}\lambda ' +\lambda '^2\right] & = & -8\pi p_n ,
\label{eeq4}
\end{eqnarray}

\begin{eqnarray}
3\dot\lambda^2+\frac{n(n-1)}{2}\dot\mu ^2 + 3n\dot\lambda\dot\mu-
e^{-2\lambda}\left[2 \lambda ''+n\mu '' +\frac{4}{r}\lambda '+n\lambda '\mu '
+\right.  & &  \nonumber \\
\left. \frac{n(n+1)}{2} \mu '^2+\frac{2n}{r}\mu ' +\lambda '^2\right] & = & 8\pi \rho
\label{eeq5}
\end{eqnarray}
\end{mathletters}
(as usual, a dot denotes derivative with respect to time, and a prime, with respect to 
the radial coordinate).

The restriction to the case $\lambda = \lambda (t)$
gives the following equations of motion:
\begin{mathletters}

\begin{equation}
-\dot\mu '+\dot\lambda \mu ' -\dot\mu \mu ' =0 ,
\label{eeq1s}
\end{equation}

\begin{equation}
2\ddot\lambda + 3\dot\lambda^2 + 2n\dot\lambda\dot\mu+n\ddot\mu+
\frac{n(n+1)}{2}\dot\mu^2-e^{-2\lambda}\left[\frac{2n}{r}\mu ' -\frac{n(1-n)}{2}\mu '^2\right]=-8\pi p_3 ,
\label{eeq2s}
\end{equation}

\begin{equation}
2\ddot\lambda + 3\dot\lambda^2 + 2n\dot\lambda\dot\mu+n\ddot\mu+
\frac{n(n+1)}{2}\dot\mu^2-e^{-2\lambda}\left[ n\mu '' + \frac{n}{r}\mu ' +
\frac{n(n+1)}{2}\mu '^2\right]=-8\pi p_3 ,
\label{eeq3s}
\end{equation}

\begin{eqnarray}
(n-1)\ddot\mu+\frac{n(n-1)}{2}\dot\mu ^2+3(n-1)\dot\lambda\dot\mu +
3\ddot\lambda+6\dot\lambda ^2 - e^{-2\lambda}\left[ (n-1)\mu '' \right. &  & \nonumber  \\
\left. +\frac{n(n-1)}{2}\mu '^2+\frac{2(n-1)}{r}\mu ' \right] & = & -8\pi p_n ,
\label{eeq4s}
\end{eqnarray}

\begin{equation}
3\dot\lambda^2+\frac{n(n-1)}{2}\dot\mu ^2 + 3n\dot\lambda\dot\mu-
e^{-2\lambda}\left[ n\mu '' +\frac{n(n+1)}{2}\mu '^2+\frac{2n}{r}\mu ' \right]=8\pi \rho .
\label{eeq5s}
\end{equation}
\end{mathletters}
In the case $n=1$, these equations reduce to the ones given in Chaterjee {\em et al} 
\cite{hin1}.

It is easy to show that Eq.(\ref{eeq1s}) can be rewritten as
\beq
\mu '' + \mu ^{'2} +\frac{1}{2}\mu ' \Phi (r)=0,
\label{con1}
\eeq
where $\Phi (r)$ is an arbitrary function. Besides, by substracting Eq.(\ref{eeq2s}) to 
Eq.({\ref{eeq3s}) we get
\beq
\mu '' + \mu '^2-\frac{1}{r}\mu '=0.
\label{con2}
\eeq
So for the last two equations to be compatible we must choose $\Phi (r) = -\frac{2}{r}$. 
Eq.(\ref{con2}) is integrable and the result is
\beq
e^{\mu (t,r)} = \beta (t) r^2 + \gamma (t),
\label{sol}
\eeq
where $\beta(t)$ and $\gamma (t)$ are arbitrary functions of time. Now from Eqs.(\ref{eeq1s})
and (\ref{sol}) we get $\dot\lambda = \frac{\dot\beta}{\beta}$,
which yields
\beq
e^{\lambda (t)} = \frac 1 b\;\beta(t),
\eeq
where $b$ is an arbitrary constant. This solution is a generalization of the one obtained by Chatterjee {\em et al} for the case $n=1$ \cite{hin1}. However, it should be emphasized that due to the presence of an additional term (proportional to $1-n$) in Eq.(4b) for the case of an arbitrary $n$, this generalization is by no means trivial.

Let us remark at this point that very little is known about the behaviour of matter at extreme conditions of density and pressure in a multidimensional spacetime. So instead of adopting any particular and arbitrary equation of state, Eqs.(\ref{eeq2s}), (\ref{eeq4s}), and (\ref{eeq5s}) shall be taken as definitions 
of $p_3(t,r)$, $p_n(t,r)$, and $\rho (t,r)$, respectively. The type of matter requires to achieve dynamical compactification in the case under consideration shall be discussed below.

The model may display several different features according to 
the explicit form of the 
functions $\beta$ and $\gamma$. In the following a particular expression for these functions will be chosen, but first 
certain quantities that will be of interest in the subsequent analysis are listed: the 
scalar curvature of the $3+n$ space, the Kretschmann scalar, the expansion
scalar, and the shear scalar.

\begin{equation}
R^{(3+n)} = -9\ddot\lambda -21{\dot\lambda}^2-3n\ddot\mu-n(2n+1){\dot\mu}^2-12n\dot\lambda\dot\mu
+ne^{-2\lambda} \left[ 2\mu '' + (1+n) {\mu'}^2+\frac{4}{r}\mu '\right] ,
\label{ricci3+n}
\end{equation}

\begin{eqnarray}
K &  = &  24\dot\lambda^4+24\ddot\lambda\dot\lambda^2+12n\dot\lambda^2\dot\mu^2
+2n(n+1)\dot\mu^4+12\ddot\lambda^2+4n\ddot\mu^2+8n\ddot\mu\dot\mu^2 
+e^{-2\lambda}\left[-8n\dot\lambda^2\mu'^2 + \right. \nonumber \\  
& & \left. 16n\dot\mu'\dot\lambda\mu '-16n\dot\mu '
\dot\mu\mu '-\frac{16n}{r}\dot\lambda\dot\mu\mu ' - 8n\dot\mu '^2 
-4n(n+1)\dot\mu^2\mu '^2+ 
8n\mu '^2\dot\lambda\mu '-8n\mu ''\dot\lambda\dot\mu
\right] +\nonumber \\
& &    e^{-4\lambda}\left[ \frac{8n}{r^2}\mu '^2+2n(n+1)\mu '^4 + 4n\mu ^{2''} 
+ 8n\mu '' \mu '^2\right] ,
\label{kret}
\end{eqnarray}

\begin{equation}
\theta = 3\dot\lambda + n\dot\mu ,
\eeq

\begin{equation}
\sigma^2 = \sigma_{\mu\nu}\sigma^{\mu\nu} = (n+4){\dot\lambda}^2+n\frac{n(n+1)+9}{9}{\dot\mu}^2+
2\frac{n(n+1)}{3}\dot\mu\dot\lambda .
\label{shear}
\end{equation}
The scalar curvature of the $3+n$ space was calculated by means of the expression
\cite{raycha}

\begin{equation}
R^\mu_{\;\;\mu} = R^{(3+n)}+\dot\theta+\theta ^2 - 2\omega^2 - \dot v^\mu_{\;\; ;\mu}
\end{equation}
(It can be seen that in this model the last two terms of this expression are null).

We would like the model to describe a 4-dimensional ``macroscopic'' expanding spacetime plus $n$ ``microscopic'' dimensions. We must impose 
in consequence 
the
conditions $\dot\lambda >0$, and $\dot\mu <0$ on the metric functions (the latter must be 
valid at least for some part of the 
evolution of the universe). This can be easily achieved by taking advantage 
of the freedom in the
arbitrary functions $\beta$ and $\gamma$. Let us take as an example
\beq
\beta (t) = -b \ln (1+t^\alpha ),\;\;\;\;\;\;\;\;\;\;\;\;\;\;\;\;\;\;\;\;\;
\gamma (t) = a \ln (1+t^\alpha )-k t^\delta .
\eeq
With this choice, the functions appearing in the metric are 
\beq
e^\lambda = \ln (1 + t^\alpha ), \;\;\;\;\;\;\;\;\;\;\;\;\;\;\;\; e^\mu = (a-br^2)\ln (1+t^\alpha ) - k t^\delta
\eeq
Note that the constant $b$ is a measure of the inhomogeneity of the model. 
The positively defined and arbitrary constants $a$, $b$, $k$, 
$\alpha$, and $\delta$ are to be 
chosen in a convenient way. 

Next, some comments on the behaviour of the scale factors. Due to the election made for $\beta (t)$ and $\gamma (t)$, the big bang is synchronous for both scales. Besides, in order that $\dot\mu$ be negative from some
point of the evolution onwards, we have to demand that $a-br^2>0$. In this case, the scale factor of the 
three space is monotonically increasing, while
the scale factor of the internal dimensions grows until a time $t_{\rm max}$, given for each $r$
by 

\beq
r^2 = \frac{a}{b}-\frac{k\delta}{b\alpha}\;t_{\rm max}^{\delta -\alpha}\;(1+t_{\rm max}^\alpha),
\label{tmax}
\eeq
and compactifies dynamically to zero size 
at different times $t_0$ for each $r$, given by

\beq
r^2= \frac{a\ln (1+t_0^\alpha)-kt_0^\delta}{b\ln (1+t_0^\alpha)}.
\label{tcero}
\eeq

We move now to the analysis of the asymptotic behaviour 
of the model. From the explicit expression of $K$ and 
$R$ it is seen that both of them diverge as $t\rightarrow 0$, the first one as $t^{-4}$, and
the second one as $t^{-2}$. The existence of the initial singularity is confirmed by 
the divergence at $t=0$ of $\rho ,p_3$ and $p_n$. At $t=t_0$, all the matter functions, the curvature 
scalars and the shear scalar diverge. It has been argued however \cite{sahdev}
that there might exist some sort of
stabilization mechanism (probably due to quantum gravity 
effects) which could prevent the formation of
the final singularity \cite{appel}. This would allow the evolution of
the ordinary 3-space independently of the internal (and microscopic) space, as can be seen from the equation of the conservation of the energy. However, one must be careful at this point.  
It must be emphasized that no matter which the stabilization mechanism is, the resulting function $\mu$, along with $\lambda$, must still be a physically sensible solution of the $4+n$ dimensional equations of motion after the compactification (see \cite{bl} and \cite{sahdev}). For instance, if we assume that $\mu = \mu_0$ = constant after the compactification \cite{sahdev}, then the pressure in the internal space must be constant in the post-compactification phase. However, this contradicts Eq.(4d) which implies than $p_n$ is a function of $t$ if $\mu = \mu_0$. Obviously, the ultimate stabilization mechanism (if any) will be determined by the still elusive quantum theory of gravitation. In the meantime, any claim about the post-compactification phase of the system must be in agreement with the physical consequences of the stabilization mechanism chosen. This fact has been frequently overlooked in the literature on this subject \cite{ban20}.

Finally we analyze the behaviour of the fluid in the light of the strong and weak energy conditions. In the case of SEC, the quantity of interest is 
\beq
R_{\mu\nu} v^\mu v^\nu = \frac{8\pi G}{2+n}[ (1+n)\rho + 3p_3 + np_n] ,
\label{sec}
\eeq
where $R_{\mu\nu}$ is the $4+n$-dimensional Ricci tensor and $v^\mu$ is the velocity of the fluid \footnote{ The expression (\ref{sec}) plays an important role in the problem of singularities in higher-dimensional spacetimes. See \cite{ban21}.}. To simplify the calculations, we adopt the particular case in which $\alpha = \delta = 1$. In this case, 

\beq
R_{\mu\nu}v^\mu v^\nu  = \frac{ (n+3) (a-br^2)\ln (1+t) -3kt}{(1+t)^2\ln (1+t)}\;e^{-\mu}.
\eeq
The fact that this expression is positive for all the values of $t$ in the interval $(0,t_0)$ implies that the matter satsfies SEC for all $t$ and $r$ in the $4+n$-dimensional phase.

In the case of WEC, the important quantity is the matter density, given by
\beq
\rho (r,t) = \frac{[R_1(t)\;r^4 + R_2(t) \;r^2 + R_3(t)]\ln (1+t) + 6k^2t^2}{4(1+t)^2 \ln (1+t)}\;e^{-2\mu} ,
\label{dens}
\eeq
where
\beq
R_1(t) = 2b^2(n^2 +5n +6) ,
\nonumber
\eeq
\begin{eqnarray}
R_2(t) &  =  & -8nb^2(n+2)\;t^2 + 2b[2n^2(k-4b)+n(7k-16b)+6k]\;t + \\ \nonumber  & & 4b[n^2(k-a-2b)+2n(k-5a-2b)-6a],
\end{eqnarray}
\begin{eqnarray}
R_3(t)  & = & -12nkbt^3 + 2n[12b(a-k)+k^2(n+2)]\;t^2 + 2[2n^2k(k-a)+n(24ab+k^2- \\ \nonumber
& & 7ka-6bk)-6ka]\;t +2[n^2(a-k)^2 + n (12ab + 5 a^2 - 4ka - k^2) + 6a^2].
\end{eqnarray}
To completely avoid WEC violation, the numerator of Eq.(\ref{dens}) must be positive for all values of the variables $r$ and $t$. The fulfillment of this conditions depends crucially on the values of the constants $a,b$ and $k$. However, we expect that for a given $t$, the $3+n$-dimensional space could be resolved in two types of regions, according to whether the matter in each region satisfies WEC or not. It follows from Eq.(\ref{dens}) that the distribution of these regions will be inhomogeneous.

\section{Conclusions}

It was shown that there exists a family of solutions, parameterized by the 
functions $\beta$ and $\gamma$, for the very complex system of equations
corresponding to the case of a $4+n$-dimensional inhomogeneous model. The matter content of the model satisfies, in  the pre-compactification phase, the SEC for every value of $t$ and $r$, and the WEC in some regions of spacetime. A general feature of these solutions 
is that the time at which dynamical compactification of the extra dimensions 
begins is different for each value of the $r$ coordinate. The particular example that was analyzed here evolves from
a $4+n$-dimensional into a 4-dimensional spacetime, the features of which depend on the stabilization method. 
It is worth pointing out again that any claim about the evolution after the compactification must be consistent with the adopted compactification
scheme and with the higher-dimensional equations of motion.

{\bf Acknowledgements:} The author would like to thank CLAF-CNPq for financial support, 
and J. Salim, M. Novello, and A. Krasi\'nski  for helpful comments.

\clearpage

\end{document}